\documentclass[preprint,apl,notitlepage]{revtex4-1}

\usepackage{graphicx}
\usepackage{amssymb}
\usepackage{amsfonts}
\usepackage{amsmath}
\usepackage{amsbsy}
\usepackage{natbib}
\usepackage{comment}
\usepackage{color}
\usepackage[dvipsnames]{xcolor}
\graphicspath{{./}}
\usepackage{subcaption}
\usepackage{hyperref}
\usepackage{cleveref}
\usepackage{changes}
\usepackage{epstopdf}
\renewcommand{\vec}[1]{\mbox{\boldmath$\mathrm{#1}$}}
\def\ind#1{{_{\mathrm{#1}}}}
\newcommand{\be}{\begin{equation}}
\newcommand{\ee}{\end{equation}}
\newcommand{\ben}{\begin{eqnarray}}
\newcommand{\een}{\end{eqnarray}}
\newcommand{\dd}{\mathrm{d}}

\begin{document}

\title{Ultrafast imprinting of topologically protected magnetic textures via pulsed electrons}
\author{A.~F.~Sch\"{a}ffer$^1$, H.~A.~D\"urr$^2$, J.~Berakdar$^1$}
\affiliation{$^1$Institut f\"ur Physik, Martin-Luther-Universit\"at Halle-Wittenberg, 06099 Halle (Saale), Germany}
\affiliation{$^2$Stanford Institute for Materials and Energy Sciences, SLAC National Accelerator Laboratory, 2575 Sand Hill Road, Menlo Park, California 94025, USA}
\begin{abstract}
Short electron  pulses are demonstrated to trigger and control magnetic excitations, even at low electron current densities.
 We show that the tangential magnetic field surrounding a picosecond electron pulse can imprint topologically protected magnetic textures such as skyrmions in a sample with a residual Dzyaloshinskii-Moriya  spin-orbital coupling. Characteristics  of the created excitations such as the topological charge can be steered via the duration and the strength of the electron pulses.
The study points to a possible way for a spatio-temporally controlled generation of skyrmionic excitations.
\end{abstract}

\date{\today}

\maketitle

\begin{figure}[!th]
				\includegraphics[width=.6\linewidth]{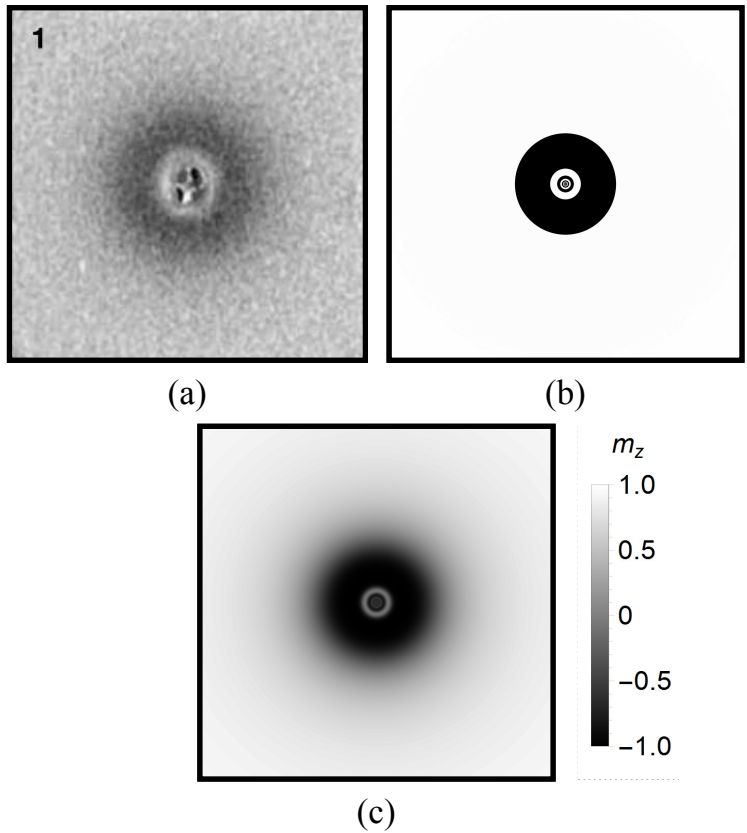}
				\caption{Comparison between experimental (a)\cite{27},   and numerical results (b), (c).
				Both numerical simulations and the experimental data cover an area of $150\times 150\,\mu$m$^2$. In contrast to panel (b),  in (c) the demagnetizing fields are included in simulations.
				The grey shading  signals the magnetization's $z$-component with white color  meaning $m_z=+\hat{e}_z$ and black  $m_z=-\hat{e}_z$. The electrons in the
				beam impinging normal to the sample have an energy of 28\,GeV. The pulse's time-envelope is taken as a Gaussian  with a pulse duration of $\sigma_t = 2.3\,$ps, which translates to a number of $n_e\approx  10^{10}$ electrons and an equivalent  time-dependence of the generated Oersted field whose radial $\rho$ dependence away from the beam axis derives to  $B(\rho)=54.7\,$T$\mu$m$/(\rho+\epsilon)$ (at the peak electron bunch intensity). The cut-off distance $\epsilon=40\,$nm is included in order to avoid a divergent behavior at the origin and can be understood as a rough approximation of the beam width.
				 }
				\label{fig_comp}
		\end{figure}

\begin{figure}[!th]
			\includegraphics[width=0.6\linewidth]{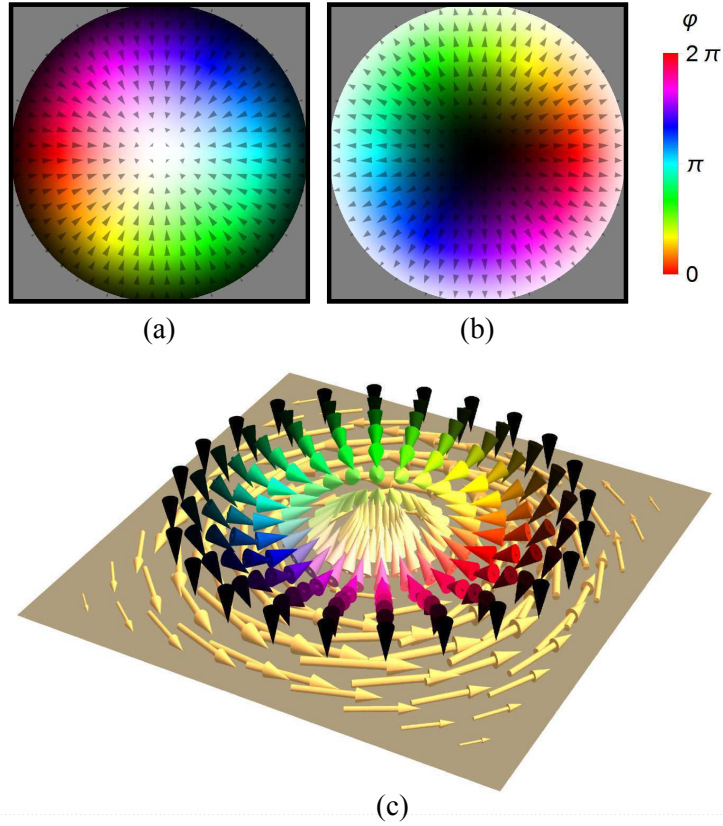}
			\caption{Magnetic ground states for a nano disc with a diameter of 300\,nm and a thickness of 1.5nm. The material parameters are $M\ind{sat}=450\times 10^3\,$A/m, $A\ind{ex}=10\,$pJ/m, $\alpha=0.01$, $K_u=1.2\times 10^5\,$J/m$^3$ (out-of plane anisotropy), and the interfacial DMI-constant $D\ind{ind}=0.31\times 10^{-3}\,$mJ/m$^2$. (a) corresponds to $N=1$, whereas (b) possesses $N=-1$, both skyrmions are of the N\'eel type.  Bottom panel illustrates pictorially  the influence of the magnetic field associated with the
					electron bunch. The cones correspond to the initial magnetic configuration as in (a) and (b), whereas the golden arrows show the induced magnetic field. The resulting torque points perpendicular to the magnetization, affecting the magnetic configuration accordingly. }
			\label{fig_groundstate}
		\end{figure}

Tremendous progress has been made towards the realization of spatiotemporally controlled electron sources for probing the materials local structural, electronic and magnetic dynamics \cite{1,2,3,4}.
Working schemes rely on the electron emission from a laser-irradiated nanoscale apex \cite{5,6,7,8,9,10,11,12,13,14,15,16,17,18,19,20,21,22} with the
 electron pulse duration being controllable  with the 
laser pulse duration. The   laser intensity dictates  the electron number   in the bunch.  
Electron pulse acceleration and control is achievable by   
intense THz)fields \cite{23,24,25}.
Here we explore the potential 
of very fast, relativistic  electron bunches for a possible  control of the magnetic dynamics in a thin film which is traversed by the electrons. 
Our focus  is on the sample spin dynamics    triggered by
the  electric and magnetic fields  associated with the electron bunch \cite{26}.
In fact,  a pioneering experiment \cite{27} explored the ultimate speed limit for precessional  magnetic dynamics of CoCrPt film driven by the magnetic field  $\vec{B}(\vec{r},t)$ of
short relativistic electron pulses (with a duration of $\delta = 2.3\,$ps) passing a 14nm thin film of granular CoCrPt ferromagentic samples with grain sizes of $20.6\pm 4$nm.
The main experimental results are shown along with our simulations  in Fig.\ref{fig_comp}.
Prior to the electron-pulse the sample was magnetized homogeneously in $z$ direction. The pulse induced ring pattern of the magnetic domains pointing either up or down (with respect to  the easy direction of the magnetic films) is well captured by our micromagnetic simulations and can  be interpreted by the analytical model enclosed  in the supplementary materials.
As pointed out in \cite{27},  the critical precessional angle $\phi\geq\pi/2$ is determined  by the local strength of the magnetic field and indicates  the achieved angular velocity $\omega$. 
The pulse duration $\delta$ plays a crucial role \cite{28}.  As discussed in Ref.\cite{28}, an appropriate sequence of ps pulses allows for  an optimal control scheme achieving  a ballistic magnetic switching, even in the presence of high thermal fluctuations.  Longer pulses might drive the system back to the initial state \cite{28}. So, the critical precessional angle and $\delta$ are the two key parameters  \cite{27} for the established   final precessional angle $\phi=\omega\delta$.  Note,  the demagnetization fields are also relevant, as inferred from Fig. \ref{fig_comp} but they do not change the main picture (further details are in the supplementary materials).
		Having substantiated our methods against experiment we turn to the main focus of our study, namely the generation of topologically  protected magnetic excitations such as skyrmions via the electron pulses. We consider samples exhibiting Dzyaloshinskii-Moriya (DM) spin-orbital coupling are appropriate.
		A recent work \cite{29} evidences that  ultrathin nano discs of materials such Co$_{70.5}$Fe$_{4.5}$Si$_{15}$B$_{10} $\cite{30} sandwiched between Pt  and Ru/Ta  are  well suited for our purpose.
		The magnetization's structure may  nucleate spontaneously into skyrmionic configurations.
		We adapted the  experimentally verified parameters for this sample and present here the result  for the magnetic dynamics  triggered by short electron beam pulses.
		Taking a nano disc of a variable size the ground state with a topological number $|N|=1$ is realized after propagating an initially  homogeneous magnetization in $\pm z$ direction according to the Landau-Lifshitz-Gilbert equation (LLG) including DM interactions.
		The two possible ground states, depending on the initial magnetization's direction are shown in \cref{fig_groundstate} along with the material's parameters.\\
		Our main focus is on how to efficiently and swiftly create skyrmions, an issue of relevance when it comes to practical applications.
		Previous theoretical predictions (e.g. \cite{31})  utilize a spin-polarized current for the skyrmion generation. Large currents densities and a finite spin polarization of injected currents are needed, however. Thus, it is of interest to investigate the  creation and annihilation of skyrmions with current pulses similar to those discussed above using the surrounding magnetic field. 
		%
	 Of interest is the skyrmion generation and modification via a nano-focussed relativistic electron pulse. While currently such pulses can be generated with micron size beam dimensions \cite{ued_ref} future sources are expected to reach ficus sizes down to the few nm range \cite{32}. In principle the possibility of beam damage occurring in the beam’s focus as in the case of the experiment in ref.\cite{27} is present. However, ongoing experiments with relativistic electron beams \cite{ued_ref} indicate that the use of ultra thin freestanding films may alleviate damage concerns. \\
		
		Topologically protected magnetic configurations, like magnetic skyrmions, are well defined quasiparticles. They can be characterized mathematically by the topological or winding number $N=\frac{1}{4\pi}\int \vec{m}\cdot\left(\frac{\partial \vec{m}}{\partial x}\times\frac{\partial \vec{m}}{\partial y}\right)\dd x\dd y$\cite{33} which counts simply  how often the unit vector of the magnetization wraps the unit sphere when integrated over the two-dimensional sample.
		Therefore, skyrmions are typically a quasiparticle in thin (mono)layers.
		The topological number  adopts integer values indicating the magnetic configuration to be skyrmionic ($N=\pm 1$) or skyrmion multiplexes ($|N| >1$). If the topological number is not an integer the topological protection is lifted and the magnetic texture is unstable upon  small perturbations.
		The topological stability of skyrmionic states stem from the necessity of flipping at least one single magnetic moment by $180^\circ$, to overcome the barrier and transfer the state into a "trivial" state, like a single domain or vortex state.
		In the following, we will attempt to overcome this energy barrier with the previous methods so that the magnetization will be converted into a state with a different topological invariant.
		Advantageous is the spatial structure of the magnetic field curling around the beam's center, which gives a good point of action in order to manipulate topologically protected configurations.\\

\begin{figure}[th]
			\includegraphics[width=0.6\linewidth]{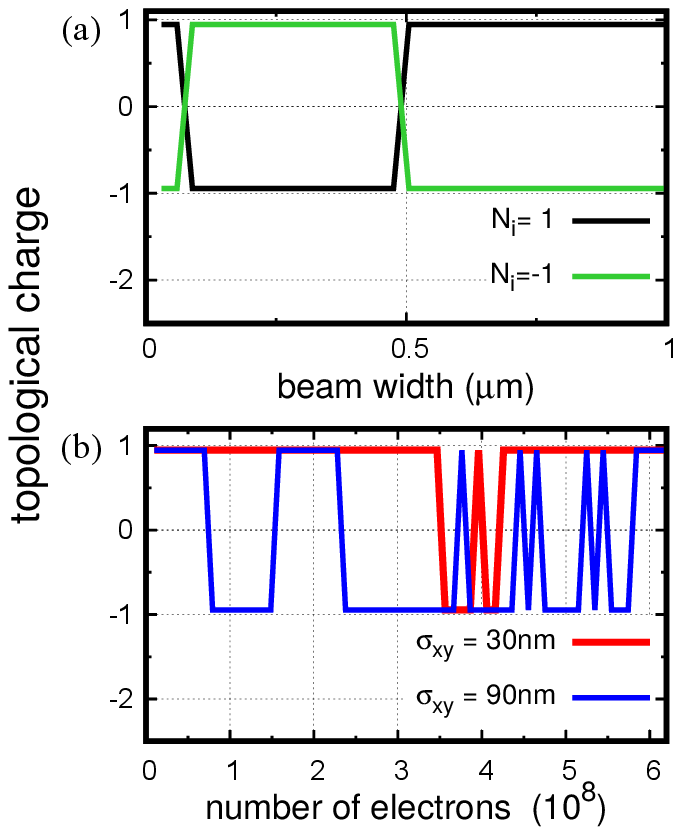}
			\caption{Varying the number of electrons per pulse or the spatial enlargement of the pulse, the imprinted  topological charge  can be tuned. The pulse duration is set to $2.3\,$ps. Black and green curves correspond respectively to starting with a magnetic ordering having $+1$ or $-1$ topological charge, as shown in \cref{fig_groundstate} for different pulse widths. Both the blue and red curve start from $N_i=+1$.
			The sample is a magnetic disc (diameter $d=300\,$nm) which is  irradiated  with a Gaussian beam pulse with $\sigma_{xy}=30\,$nm (and $90\,$nm) in case of the bottom graph, respectively the upper graphs' beam has a constant number of $n_e=10^8$ electrons. }
			\label{fig_dur}
		\end{figure}

	Using the short electron pulses one may overcome the topological energy barrier with a magnetic "kick" and the magnetization relaxes afterwards, possessing a different winding number.
	In contrast to the mesoscopic system studied in \cref{fig_comp}, in the following not only the far field, but also the near magnetic field of the Gaussian pulses will be treated. To do this the Biot-Savart law is solved numerically and fitted with a model function. For magnetic systems a minimum time of exposure is necessary, whereas the spatial focus of the beam is limited. To overcome this conflict, the pulse duration is fixed at $2.3\,$ps as before, when nothing different is mentioned. Details on the resulting magnetic field can be seen in the supplementary material. Starting from such an electron beam two main parameters can be adjusted to achieve the favored reaction of the nanodiscs. Those are the pulse width and the number of electrons, which will be treated independently. 	In \cref{fig_dur}, the final topological charges after a single Gaussian electron pulse irradiating a nanodisc are plotted as a function of the  number of electrons and the width of the Gaussian distributed electrons. 
	The results do not show the transient  time evolution of the sample but only the final steady-state  values of the winding number. They are obtained by applying an electron pulse, propagating the magnetization during the pulse, and relaxing the magnetic configuration afterwards as to  approach a local minimum of the free energy's hypersurface.
		
		We note  the strong correlation between the change of the topological charge and the number of electrons or accordingly the beam width. Relatively large intervals of both parameters lead to the same final values for $N$. We note that not only the variation of these control parameters, but also of the duration of the pulse is experimentally accessible, particularly in a nanoapex ultrafast transmission electron microscope.
		Noteworthy,  the graphs for opposite initial configurations (see \cref{fig_dur}(a)) are axially symmetric with respect to the $x$ axis. This can be explained by the coinciding symmetry centers of the pulse and the skyrmionic structure.
		This symmetric and robust behavior can be exploited to switch between the accessible different values for the topological charges which are quite close to the ideal integer values that would be realized in an infinitely small discretization. \\
		Interestingly, the switching between the two stable states occurs repetitively for an increase in the number of electrons, whereas the spatial manipulation of the beam leads to one regime only in which the fields are sufficient to switch the topological number. The first observation can be explained with the schematics shown in fig.\ref{fig_groundstate}c). Depending on the strength of the pulse the magnetic moments are forced to rotate multiple times around the $\hat{e}_\varphi$ vector in a collective manner, as each moment of equal distance to the center experiences the same torque. The final position of the surrounding moments couples backwards to the center and determines the new topological charge. The electron number linearly translates to the peak magnetic field, wheras the beam width has a more complicated influence. 
		When the width is inceased the spatial profile in the $xy$-plane is manipulated, as the maximum magnetic field is shifted towards the disc's rim and beyond. How the system reacts on this changes depends crucially on the exact profile of the beam, especially on the point of maximum magnetic field strength, as can be seen in \cref{fig_dur}(a).
		This leads to the question of the optimum parameter regime, to manipulate the system reliably, which can not finally be answered as it strongly depends on the experimentally available capabilities. Hence this work focuses on an examplary study on the effect.
		\\
		The same switching phenomenon as discussed before can also be observed for different setups. Weaker pulses, as long as they are able to overcome the internal fields to excite the system, can be used as well, but obviously the field's amplitude translates to the strength of the resulting torque. This implies a longer radiation time needed for pulses of lower intensity to be capable to switch the system.		
		In case of different materials or geometries the accessible topological states have to be investigated, before they can be utilized. Otherwise undesired, interstitial states might be achieved by accident and the switching is not deterministic anymore. 
		\\
		
\begin{figure}[th]
			\includegraphics[width=0.5\linewidth]{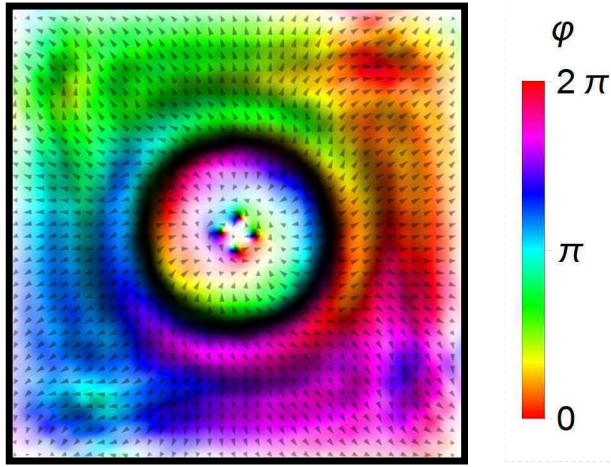}
			\caption{Magnetization configuration 1ns after a pulse with a duration of 22ps and correspondingly $10^{11}$ electrons focused on $\sigma_{xy}=10\,$nm. Four skyrmions in the inside of a domain-wall ring are protected against fluctuations outside the ring. The sample covers an area of $800\times 800\,$nm$^2$."(Multimedia view)"}
			\label{fig_ringSk}
		\end{figure}

		Aside from the manipulation of the topological charge of a given nano disc system, the creation of skyrmions on extended thin layers is an open challenge. To treat this, we start from a quadratic region with a size of $(800\times 800)\,$nm$^2$  and periodic boundary conditions in $x$ and $y$ direction to avoid finite size effects.
		To overcome the homogeneously magnetized state, the peak intensity has to be increased further, as well as a focussing down to the scale of the desired skyrmion.  The Gaussian profile in the $xy$-plane has a standard deviation of $30\,$nm, whereas the pulse duration is reduced to $500\,$fs and a single pulse includes $10^8$ electrons.
		Even though this is experimentally challenging, it is necessary to create skyrmions on an extended film. If the beam size is too large, only trivial domain rings are built up. On the other hand the duration has to be long enough to allow the magnetic texture to react to the pulse.	
		A well known characteristics for magnetic skyrmions is a tendency to  a blow-up behavior to minimize the exchange energy, when no external stabilization is present. In nano discs the stabilizing factor is the geometric confinement. Several experimental works incorporate an additional external magnetic field perpendicular to the surface to amplify the uniaxial anisotropy in an attempt to block the blow-up effect. As we are mainly interested in the generation of topological defects, we focus on this aspect keeping in mind that a stabilization of the induced skyrmions is necessary to maintain their localized structure.
		\begin{figure}[th]
			\includegraphics[width=0.6\linewidth]{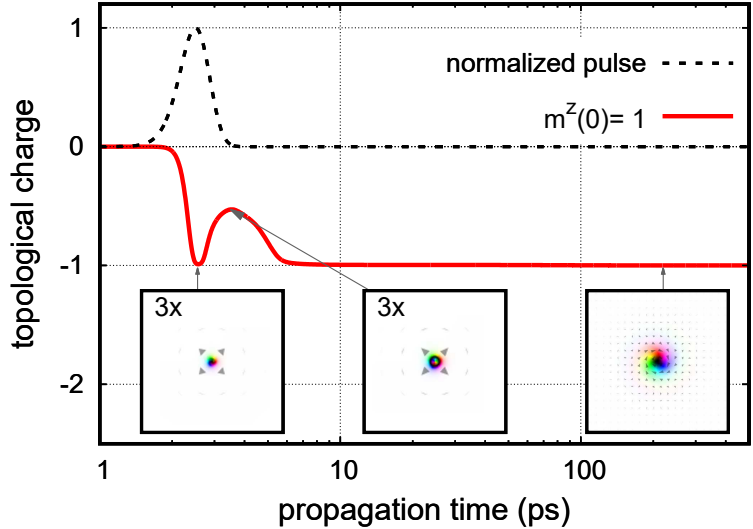}
			\caption{Evolution of the topological charge after a single Gaussian beam with $n_e=10^8$ electrons, $\sigma_{xy}=30\,$ nm and $\sigma_t=0.5\,$ps. The initial magnetization points homogeneously in $z$ direction. The system  size is $(800\times 800)$nm$^2$ with  periodic boundary conditions. On the  bottom we present a few snap-shots corresponding to the red curve  at time moments  depicted on each snapshot. The first and second picture is scaled with a factor of three in order to make the development clearer.}
			\label{fig_EvHom}
		\end{figure}
		As can be seen in \cref{fig_EvHom}, the irradiation by the electron beam leads to the injection of a topological charge. The used pulse has the peak intensity at $5\sigma_t=2.5\,$ps to account for its finite rising time. After short oscillations the system relaxes to a skyrmionic state. On the time scale of the electron pulse, the state's topological nature is perfectly   stable but the excitations tend to expand when no further stabilizations are present.
		Therefore, the diameter of the skyrmion increases and the type changes from N\'eel to Bloch, the topological charge is still conserved. 
		The situation changes markedly  for longer pulses ($\sigma_t \gtrsim 20\,$ps).
		Just like before the pulse leads to topological excitations. On top of this, a domain-wall ring similar to the results presented above is induced  which shields the included skyrmions in two different ways (see \cref{fig_ringSk}).
		The blow-up behavior is blocked due to the domain wall. The spin waves reflected from the open boundaries are absorbed by the domain ring so that the inside is not affected. In the supplementary  
		\href{./animation.avi}{animation}"(Multimedia view)" the formation of the structure in \cref{fig_ringSk} can be seen. In the first 220\,ps the time step between the single frames is 4.4\,ps. Afterwards the movie is accelerated to 22\,ps/frame which can be easily recognized by the arising spin-waves outside the domain ring.
		Applying another pulse the ring structure can be opened and the topological charge changed with the same process.

Summarizing, we demonstrated the  usefulness of  ultrashort electron pulses for generating and steering the magnetization dynamics  due to the
	  electromagnetic fields associated with the electron pulses. In particular, topologically protected magnetic textures such as skyrmions can be imprinted and  manipulated in controllable spatiotemporal way.
\section*{Supplementary Material}
See supplementary material for information on the analytical macrospin approach towards the results shown in \cref{fig_comp}, further details on the numerical calculations and the magnetic fields induced by ultrafast electron pulses.

\section*{Acknowledgements}
A. F. S. and J. B. are supported by the German Research Foundation (Nos. SFB 762) and the Priority Programme 1840. H.A.D. acknowledges support by the U.S. Department of Energy,  Office of Basic Energy Sciences, Materials Sciences and Engineering Division under Contract No. DE-AC02-76SF00515.
%
%


\end{document}